\documentclass[useAMS,usenatbib]{mn2e}
\usepackage{txfonts}
\usepackage{graphicx}
\usepackage{pdflscape}
\usepackage{threeparttable}
\def\NH2{N$_{\mathrm{H}_{2}}$}
\def\mum{$\,\mu$m}

\def\sun{$_{\odot}$}

\def\n2h{N$_{2}$H$^{+}$}
\def\d2n{N$_{2}$D$^{+}$}
\def\nh3{NH$_{3}$}
\def\hco{HCO$^{+}$}

\def\avir{$\alpha_{vir}$}
\def\afor{$\alpha_{vir}$}
\def\ak{$a_{\mathrm{k}}$}
\def\aG{$a_{\mathrm{G}}$}

\title{Massive 70\mum\ quiet clumps II: non-thermal motions driven by gravity in massive star formation?} 

\author[A. Traficante, G. A. Fuller, et
al.]{A.~Traficante$^{1,2}$, 
  G.~A. Fuller$^{1}$, R.~J. Smith$^{1}$, N.~Billot$^{3}$, A.~Duarte-Cabral$^{4}$, N.~Peretto$^{4}$, 
\newauthor
S.~Molinari$^{2}$ and J.~E. Pineda$^{5}$ \\
$^{1}$Jodrell Bank Centre for Astrophysics, School of Physics and Astronomy, University of Manchester, Oxford Road, Manchester M13 9PL, UK \\
$^{2}$IAPS - INAF, via Fosso del Cavaliere, 100, I-00133 Roma, Italy \\	
$^3$ Instituto de Radio Astronom\` ia Milim\` etrica, Granada, Spain\\
	$^{4}$School of Physics and Astronomy, Cardiff University, Queens Buildings, The Parade, Cardiff CF24 3AA, UK\\
	$^5$ Max-Planck-Institut fur extraterrestrische Physik (MPE), Giebenbachstrasse 1, D-85741 Garching, Germany}

\begin{document}
\maketitle

\label{firstpage}

\begin{abstract} 
The dynamic activity in massive star forming regions prior to the formation of bright protostars is still not fully investigated. In this work we present observations of HCO$^+$ $J=1-0$ and N$_2$H$^+$ $J=1-0$ made with the IRAM 30m telescope towards a sample of 16 {\it Herschel}-identified massive 70\mum\ quiet clumps associated with infrared dark clouds. The clumps span a mass range from 300 M$_\odot$ to 2000\,M$_\odot$.  The N$_2$H$^+$ data show that the regions have significant non-thermal motions with velocity dispersion between 0.28 km s$^{-1}$ and 1.5 km s$^{-1}$, corresponding to Mach numbers between 2.6 and 11.5. The majority of the 70\mum\ quiet clumps have asymmetric HCO$^+$ line profiles, indicative of significant dynamical activity. We show that there is a correlation between the degree of line asymmetry and the surface density $\Sigma$ of the clumps, with clumps of $\Sigma\gtrsim0.1$ g cm$^{-2}$ having more asymmetric line profiles, and so are more dynamically active, than clumps with lower $\Sigma$. We explore the relationship between velocity dispersion, radius and $\Sigma$ and show how it can be interpreted as a relationship between an acceleration generated by the gravitational field \aG, and the measured kinetic acceleration, \ak, consistent with the majority of the non-thermal motions originating from self-gravity. Finally, we consider the role of external pressure and magnetic fields in the interplay of forces. 

\end{abstract}

\section{Introduction}\label{sec:intro}
Pioneering studies of star-forming regions showed a significant contribution from the ambient interstellar medium (ISM) turbulence in dictating the properties of clouds. This was first summarized in the Larson's line-width size relation $\delta v\propto\mathrm{R}^\Gamma$, with $\Gamma=0.38$ \citep{Larson81}. Later this value was modified to $\Gamma=0.5$, consistent with early observations of giant molecular clouds (GMCs) which show that GMCs have all similar mass surface densities, $\Sigma$, and turbulence is ubiquitous in the ISM \citep[ e.g.][]{Solomon87,Heyer04}. The exponent $\Gamma=0.5$ is expected if the interstellar medium is modeled as a turbulent fluid dominated by shocks \citep[][ and references therein]{McKee07} and simulations confirm that GMCs and their embedded massive clumps are shaped by supersonic motions and follow the Larson's relationship with $\Gamma=0.5$ \citep[e.g.][]{Padoan02,Field08}.

More recently the study of \citet[][hereafter H09]{Heyer09} questioned the Larson relation and showed that the quantity $\sigma/\mathrm{R}^{0.5}$ is not a constant of the clouds. H09 used data from the Boston University-FCRAO Galactic Ring Survey of $^{13}$CO J$=1-0$ emission \citep{Jackson06} to revise the study of the GMCs done by \citet{Solomon87} using the $^{12}$CO J$=1-0$ emission. The latter is a tracer of low density gas, which rapidly becomes optically thick within the denser regions of the GMCs. The H09 analysis shows that the average mass surface density $\Sigma$ of the clouds is not constant, and that $\sigma/\mathrm{R}^{0.5}$ increases as function of $\Sigma$, which we will refer to as the `Heyer relation'.

It is unclear, however, how the Heyer relation may or may not cascade down to the scales of clumps and cores. While inside low-mass cores the turbulence seems to dissipate down to thermal levels \citep[i.e. following the standard Larson's relation $\sigma\propto$R$^{0.5}$,][]{Fuller92}, there are some indications that this does not happen in massive star-forming clumps and cores \citep{Ballesteros-Paredes11}. For instance, combining the data from H09 with a sample of massive clumps identified in IRDCs from \citet{Gibson09}, \citet{Ballesteros-Paredes11} interpreted the Heyer relation as `universal', from GMCs down to clump scales. In the \citet{Ballesteros-Paredes11} model, the observed large velocity dispersions are driven by the non-thermal motions in the collapsing clouds and cloud fragments in a hierarchical and chaotic fashion \citep[see also][]{Vazquez-Semadeni09,Ibanez-Meja15}. 

In an alternative view, the observed non-thermal motions in massive regions could be attributed to increasing internal turbulence required to maintain equilibrium and slow-down the otherwise fast collapse, as predicted by the turbulent core model of massive star formation \citep{McKee03}.

Whether the observed non-thermal motions in high-mass clumps and cores are in any way related to a surface density threshold, $\Sigma_{t}$, below which massive star formation cannot occur \citep{Kauffmann13}, or if $\Sigma_{t}$ even exists, are still open questions.

Models suggest different values for $\Sigma_{t}$ depending on the contribution of the magnetic fields, with \citet{Tan14} suggesting $\Sigma_{t}$=0.1 g cm$^{-2}$ in the presence of magnetic fields, and \citet{Krumholz07} deriving $\Sigma_{t}$=1 g cm$^{-2}$ for non-magnetised regions.
Recent surveys of massive star-forming clumps show that they span a range of surface densities, with possible $\Sigma_{t}$ values ranging from $\Sigma_{t}\cong0.05$ g cm$^{-2}$ \citep[e.g.][]{Urquhart14} to $\Sigma_{t}\cong0.1-0.2$ g cm$^{-2}$ \citep[e.g.][]{Tan14,Traficante15b}.

In this work we discuss the results from a study of 70\mum\ quiet clumps, focusing on the properties of their non-thermal motions.  Since these high surface density objects are probable precursors to massive stars but are as yet unaffected by stellar feedback, they are ideal candidates for the study of the dynamical state of the initial conditions for massive star formation. 

This paper is structured as follows: in Section \ref{sec:observations} we describe the observations of a sample of 16 clumps identified in the IRDC survey of starless and protostellar clumps of \citet{Traficante15b} which we followed up with the IRAM 30m telescope\footnote{IRAM is supported by INSU/CNRS (France), MPG (Germany) and IGN (Spain).} to trace the gas kinematics. In Section \ref{sec:results} we describe the properties of the clumps used in this work and derived from dust and gas observations (\ref{sec:physical_properties}) and we investigate the clump dynamics and the degree of line asymmetry in the \hco\ spectra  (\ref{sec:clump_dynamics}). In Section \ref{sec:gravo_turbulent description} we explore the clump stability using a classical virial analysis and using a formulation which compares the acceleration driven by different forces. In Section \ref{sec:gravitationally_driven} we explore how the relation between the gravitational acceleration and the non-thermal motions varies from GMCs to massive clumps and cores, combining our data with results from the
literature (\ref{sec:Heyer}). We also explore the effect of including an external pressure (\ref{sec:external_pressure}) and magnetic fields (\ref{sec:magnetic_field}) on the force balance. Finally, in Section \ref{sec:conclusions} we summarise our results and present our conclusions.

\section{Observations}\label{sec:observations}

In this work we focus on the properties of the non-thermal motions of a sample of 16 clumps selected to be 70\mum\ quiet, have a surface density $\Sigma\geq0.05$ g cm$^{-2}$, a mass of $300\leq\mathrm{M}\leq2000$ M\sun, and a low-luminosity mass ratio L/M$<0.3$. A comprehensive description of the clump properties estimated from the dust continuum and a more extensive analysis of the molecular line data are given in \citet[][hereafter, PI]{Traficante17_PI}.

These clumps are located at distances ranging between $3.5\leq d\leq5.8$ kpc and have been observed with the IRAM 30m telescope using \n2h (1-0), HNC (1-0) and \hco (1-0) emission lines. The observations were carried out on June 2014 under the project number 034-14 in good weather with system temperatures between 92 K and 162 K. Each source was mapped in On-The-Fly mode to cover a $2\arcmin\times2\arcmin$ wide region with typical rms noise levels of $0.13\leq\sigma\leq0.32$ K in each 0.2 km s$^{-1}$ spectral channel. The telescope pointing was checked every 2 hours and the pointing error is estimated to be $<3\arcsec$. At the frequency of these transitions the beam size of the telescope is $\simeq27\arcsec$, corresponding to $\simeq0.6$ pc at a distance of 4.8 kpc, the average distance of these clumps (PI).


\section{Results}\label{sec:results}

\subsection{Clump physical properties}\label{sec:physical_properties}
To investigate the physical properties of our clumps, we make use of the mass surface density estimated from the Hi-GAL data (PI). We also use the emission from \n2h, an optically thin tracer of quiescent gas \citep{Vasyunina11} not strongly affected by infall or outflow motions, to measure the systemic velocity of the clumps and their turbulence.

Table~\ref{tab:virial_params} presents a summary of the properties of the clumps and the observational results discussed in this work. As well as the positions of the sources, it gives the clump sizes and surface densities (PI) and the non-thermal component of the 1-D velocity dispersions of the \n2h\ emission. The table also lists the Mach number of these non-thermal motions, and the virial parameter of the clumps.

The clump radius R was determined from the 2D-Gaussian fit of the clump dust emission at 250 \mum. The clump mass is estimated with a single-temperature greybody fit of the source fluxes at 160, 250, 350 and 500 \mum\ from Hi-GAL and, when available, including fluxes at 870 \mum\ from the ATLASGAL \citep{Schuller09} maps and at 1.1 mm from the Bolocam Galactic plane survey \citep[BGPS,][]{Aguirre10} maps. The mass surface density, $\Sigma$, has been evaluated within the region defined by the clump radius (PI). 

Since all the clumps are resolved in the \n2h\ observations, the 1D velocity dispersion, $\sigma_{\mathrm{obs}}$, has been determined using the CLASS task \texttt{hsf} to fit the hyperfine structure of the \n2h\ spectra, averaged across all the pixels within the clumps. The non-thermal motions $\sigma_{\mathrm{nth}}$ are estimated as $\sigma_{\mathrm{nth}}=(\sigma_{\mathrm{obs}}^2-\sigma_{\mathrm{th}}^2)^{1/2}$, where $\sigma_{\mathrm{th}}$ is the thermal component of the velocity dispersion, $\sigma_{th}=(k_{\mathrm{B}}T_{\mathrm{kin}}/(\mu_{\mathrm{N_{2}H^{+}}} m_{\mathrm{H}}))^{1/2}$, where $k_{\mathrm{B}}$ is the Boltzmann constant, $m_{\mathrm{H}}$ the hydrogen mass and $\mu_{\mathrm{N_{2}H^{+}}}$ the mean molecular weight ($\mu_{\mathrm{N_{2}H^{+}}}=29.02$). The gas kinetic temperature T$_{\mathrm{kin}}$ is fixed for each clump and equal to 10 K, comparable to the average dust temperature of these clumps ($\sim$11.2 K, PI). This leads to $\sigma_{\mathrm{th}}\simeq0.05$\,km/s. Since the average non-thermal motions of the clumps is $\sigma_{\mathrm{nth}}\simeq 0.8$\,km/s, the contribution of the thermal velocity dispersion to $\sigma_{\mathrm{obs}}$ is minimal. Despite the fact that a temperature gradient of up to $\simeq10$\,K can be expected across starless clumps \citep{Peretto10,Wilcock12}, using T=20\,K affects our results by less than $1\%$. The Mach number $\mathcal{M}$ is evaluated as $\sigma_{\mathrm{nth,3D}}/c_{\mathrm{s}}$, with $\sigma_{\mathrm{nth,3D}}=\sqrt{3}\sigma_{\mathrm{nth}}$ and $c_{\mathrm{s}}=(k_{\mathrm{B}}T_{\mathrm{kin}}/(\mu_{H}m_{\mathrm{H}}))^{1/2}\simeq0.19$ km s$^{-1}$, with a mean molecular weight $\mu_{H}=2.33$. We find $\mathcal{M}\geq2$ for all the clumps, with an average value of $\overline{\mathcal{M}}=7.54$.

\subsection{Clump dynamics from \hco}\label{sec:clump_dynamics}
To parameterise the asymmetry of the HCO$^+$ line we calculate the
parameter $\mathcal{A}_\mathrm{TG}$ as defined by
equation~\ref{eqn:asymm} where $I(v_i$) is the intensity at velocity
$v_i$ and $G(v_i)$ is the Gaussian fit to line, evaluated at velocity
$v_i$. To restrict the comparison to the HCO$^+$ gas kinematically
associated with dense gas traced by the N$_2$H$^+$ the sum was carried
out over a velocity range corresponding to $\pm5$ times the velocity
dispersion of the N$_2$H$^+$. This multiple was estimated by comparing
the velocity widths of the N$_2$H$^+$ and HCO$^+$ emission, visual
inspection of the spatial distribution of the HCO$^+$ as a function of
velocity with the column density maps and comparison with a manual,
source-by-source selection of the velocity range.
 \begin{eqnarray}
\mathcal{A}_\mathrm{TG}&=&\frac{1}{A}\left(\sum_i\left|
I\left(v_i\right)-G\left(v_{i}\right) 
\right|-B\right)
\label{eqn:asymm}
\end{eqnarray}
The quantity $A$ is the peak intensity of the Gaussian fit to the
HCO$^+$ line profile. This factor is necessary since the
numerator of $\mathcal{A}_\mathrm{TG}$ is an integrated intensity, so
that without the
normalisation, for two lines with the same intrinsic shape, the stronger
line will have a larger value of $\mathcal{A}_\mathrm{TG}$ than the
weaker line. The term in the summation represents the residual after the subtraction of the Gaussian fit to the original spectrum. Taking the modulus of the difference to measure the line asymmetry
results in the noise in the spectrum producing a positive
value for $\mathcal{A}_\mathrm{TG}$, even for a symmetric line. To correct for
this, a value $B$, the bias, is subtracted when calculating the line
asymmetry. Since the value of $B$, the offset due to noise, depends on
both the noise level in the spectrum and the velocity range over which
the emission is detected, $B$ was estimated for each source by summing
the modulus of the noise over the same velocity width as the line asymmetry was
determined, but centered at several different velocities away from the
line.

\begin{figure}
\centering
\includegraphics[width=8cm]{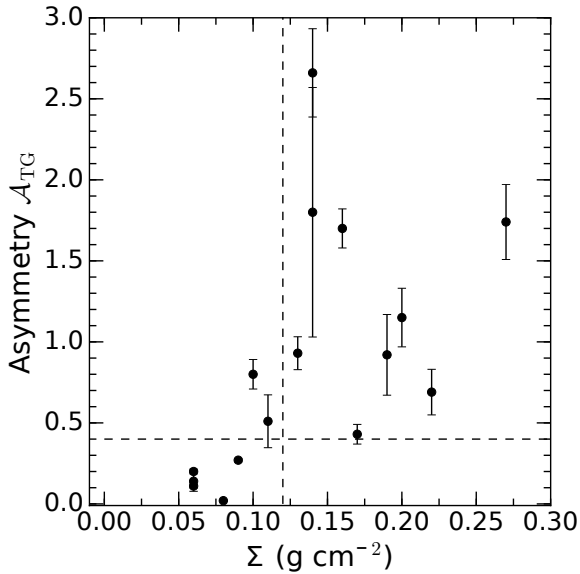}
\caption{ Asymmetry parameter, $\mathcal{A}_{\mathrm{TG}}$, as a function of clump
  surface density, $\Sigma$. The horizontal dashed line indicates the value $\mathcal{A}_{\mathrm{TG}}=0.4$ which separates symmetric from asymmetric line profiles. The vertical dashed line delimits the surface density threshold $\Sigma$=0.12 g cm$^{-2}$, the value that maximizes the difference between more dynamically active clumps and the rest of the objects, as discussed in the text.}
\label{fig:asymm}
\end{figure}

Figure~\ref{fig:asymm} shows the asymmetry parameter with its uncertainties plotted against the clump surface density, $\Sigma$. 

The value of
$\mathcal{A}_\mathrm{TG}$ for each source is given in
Table~\ref{tab:virial_params}.
Figure~\ref{fig:asymm} shows a strong correlation between the line
asymmetry and the clump surface density, with high surface density
regions having more asymmetric lines. This is confirmed by the
Pearson's correlation coefficient which gives a value of 0.55 for
$\mathcal{A}_\mathrm{TG}$ with $\Sigma$, corresponding to a
probability of $2.8\times10^{-2}$ of the correlations being due to
chance.

Clumps with $\mathcal{A}_\mathrm{TG}<0.4$ have a mean $\Sigma=0.070\pm0.006$\,g
cm$^{-2}$, while those with $\mathcal{A}_\mathrm{TG}\ge0.4$ have
$\Sigma=0.17\pm0.02$\,g cm$^{-2}$. An exact permutation test indicates
this has a probability of only $1.4\times10^{-3}$ of being due to
chance. For clumps below a surface density threshold $\Sigma_{t}=0.12$\,g cm$^{-2}$, the mean value
of $\mathcal{A}_\mathrm{TG}=0.3\pm0.1$ while clumps with
$\Sigma_{t}\gtrsim0.12$\,g cm$^{-2}$, have a mean
$\mathcal{A}_\mathrm{TG}=1.3\pm0.2$, a difference which an exact
random permutation test indicates has a probability of only
$2\times10^{-3}$ being due to chance. This threshold implies that all clumps above $\Sigma_{t}$ have $\mathcal{A}_\mathrm{TG}>0.4$. Two clumps have $\Sigma\simeq0.1$\,g cm$^{-2}$ and $\mathcal{A}_\mathrm{TG}>0.4$. Lowering the value of $\Sigma_{t}$ to include them would lead to a difference between the two sub-samples which an exact random permutation test indicates has a probability of
$4\times10^{-3}$ being due to chance, twice the probability obtained assuming $\Sigma_{t}=0.12$\,g cm$^{-2}$. Therefore, we adopt $\Sigma_{t}=0.12$\,g cm$^{-2}$ as the threshold which maximises the difference between the two group of clumps. For convenience we label (in Table~\ref{tab:virial_params}) sources with $\mathcal{A}_\mathrm{TG}\geq0.4$ as having asymmetric lines (more dynamically active) and the others as having symmetric lines. The HCO$^+$ line profiles are showed in PI.

\begin{figure}
\centering
\includegraphics[width=8cm]{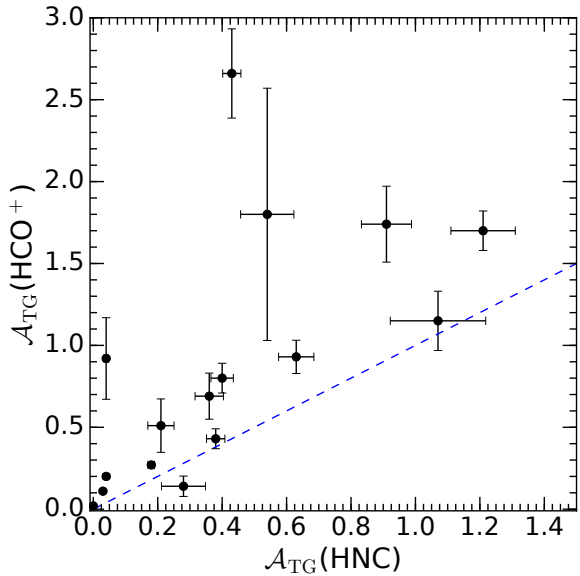}
\caption{Comparison between $\mathcal{A}_\mathrm{TG}$(\hco) and $\mathcal{A}_\mathrm{TG}$(HNC). The blue-dashed line is the y=x relation.}
\label{fig:compare_asymm}
\end{figure}

\begin{figure}
\centering
\includegraphics[width=8cm]{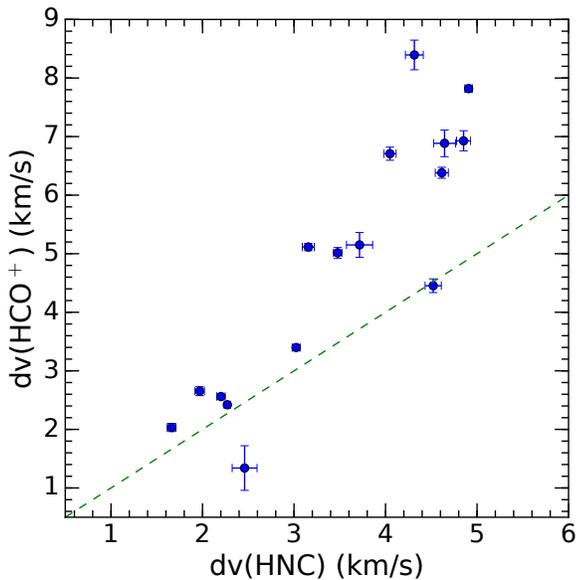}
\caption{Comparison between \hco\ and HNC linewidths. The green-dashed line is the y=x relation.}
\label{fig:compare_linewidth}
\end{figure}

The same analysis was also performed for the HNC $(1-0)$ spectra. As Figure \ref{fig:compare_asymm} shows there is a correlation between $\mathcal{A}_\mathrm{TG}$(\hco) and $\mathcal{A}_\mathrm{TG}$(HNC) with a Pearson's correlation coefficient of 0.63. In addition, the figure shows that for all but one of the sources the \hco\ line is more asymmetric than the HNC line. Comparing the $\mathcal{A}_\mathrm{TG}$(HNC) with the clump surface density shows similar results to the \hco\, but with somewhat lower statistical significance which is consistent with the evidence from Figure \ref{fig:compare_asymm} that the HNC $(1-0)$ transition is a somewhat poorer tracer of the clump dynamics. This is also consistent with the comparison of the line width of the \hco\ $(1-0)$ and HNC $(1-0)$ (Figure \ref{fig:compare_linewidth}) which shows that the \hco\ lines are typically broader than the HNC lines. A higher line broadening is a signpost of higher non-thermal motions, which may be due to a high level of turbulence and/or a significant dynamical activity, such as infall motions at the core and clump scales or outflows at the core scales. \citep[e.g.][]{Lopez-Sepulcre10,Smith13,Palau15}. \citet{Chira14} suggested that in some circumstances asymmetric line profiles in low-J transitions of optically thick lines such as \hco\ $(1-0)$ may also be interpreted as obscuration by surrounding filaments. However, the emission from \n2h\ $(1-0)$, which is optically thin in most of these sources (PI), does not show the presence of the multiple, dense gas velocity components which would be expected in the case of the emission originating from overlapping structures. Therefore, we assume that the asymmetries we observe are due to ongoing dynamical activity and the asymmetry-surface density correlation in \hco\ $(1-0)$, supported by the HNC $(1-0)$ results, show that high surface density clumps are more dynamically active that lower $\Sigma$ clumps. In particular, the lack of symmetric lines towards high surface density clumps suggests that complex dynamics are intimately connected with the presence of the highest surface density regions.

\section{A gravo-turbulent description of the star formation process}\label{sec:gravo_turbulent description}

\subsection{Virial relation and Force balance}\label{sec:virial}
\subsubsection{Classical virial analysis}

In the classical analysis of cloud stability, the non-thermal motions are due to local turbulence which support the cloud. The dynamics of the collapse is described by two independent variables, the kinetic energy $E_{\mathrm{k}}\propto\sigma^{2}$, where $\sigma$ is the 1D observed (turbulent) velocity dispersion of the region, and the gravitational potential energy $E_{\mathrm{G}}\propto$ M/R, where M is the total mass within a region of radius R. The virial parameter $\alpha_{vir}$ describes the balance between these two energies, defined as \citep{Bertoldi92}

\begin{equation}\label{eq:mass_velocity}
\alpha_{vir}=a\frac{5\sigma^2 \mathrm{R}}{G\mathrm{M}}=2a\frac{E_{\mathrm{k}}}{E_{\mathrm{G}}}
\end{equation}

\noindent where G is the gravitational constant and $a$ is a constant which includes modifications due to non-spherical and inhomogeneous density distributions. Here we assume $a=1$.

A critical value, $\alpha_{cr}$, is defined as the value of \avir\ for which the cloud is in equilibrium. If \avir\ is greater than $\alpha_{cr}$, the region will expand and dissolve, if it is lower, the region will collapse. In the absence of external pressure or magnetic fields, the hydrostatic equilibrium is reached when $E_{\mathrm{k}}$ is balanced by $E_{\mathrm{G}}$ and \avir\ has a critical value $\alpha_{cr}=1$ \citep{Tan14}. If the region is under external pressure, this pressure will work towards compressing the cloud and therefore $\alpha_{cr}$ will increase. For instance, if clouds are modeled as non-magnetized, pressure bounded isothermal spheres then they will be unstable and collapse if their mass is larger than the Bonnor-Ebert (BE) mass which leads to $\alpha_{\mathrm{cr, BE}}\cong2$ (see the discussion in \citet{Kauffmann13}; see also \citet{Tan14}). 
Conversely, in the presence of internal magnetic fields the cloud can be stabilized against collapse and therefore $\alpha_{cr}$ decreases with respect to the non-magnetized case.

The virial parameter for our sources is systematically lower than 2, and for all but one source \avir$<1$ (28.792+0.141, Table~\ref{tab:virial_params}). The average value of \avir\ is $\overline{\alpha}_{vir}\simeq0.60$.  This classical virial analysis would conclude that these clumps are all gravitationally bound. 

A virial equilibrium state is a necessary condition in the turbulent core model of massive star formation \citep{McKee03} on the assumption that in massive clumps the internal motions reflect turbulence and in absence of magnetic fields \citep{Tan14}.

Rather than being in a state of virial equilibrium it is possible that in massive regions collapse occurs in a hierarchical, global fashion which itself generates non-thermal, gravo-turbulent motion due to the chaotic collapse \citep[][]{Ballesteros-Paredes11}. In the simplest case with no external pressure or magnetic fields, the observed non-thermal motions would then arise from \textit{both} local turbulence \textit{and} self-gravity and it is \textit{not} independent from the gravitational term. This can be shown with a formulation of the problem in which the dynamics is described in terms of accelerations.

\subsubsection{Describing the Virial relation as accelerations}\label{sec:virial_accelerations}
The quantity $a_{\mathrm{G}}=\pi G\Sigma/5$ describes the average gravitational acceleration of a region, and it is a function of the mass surface density $\Sigma=\mathrm{M}/(\pi\mathrm{R}^{2})$ only. The kinetic term of the system is described by $a_{k}=\sigma^{2}/\mathrm{R}$ which also has the dimensions of an acceleration and is interpreted as the magnitude of the acceleration due to the total (thermal and non-thermal) motions in the region. If gravitational collapse significantly contributes to the observed linewidth, this would produce a correlation between $a_{\mathrm{G}}$ and $a_{k}$.

\begin{figure}
\centering
\includegraphics[width=8cm]{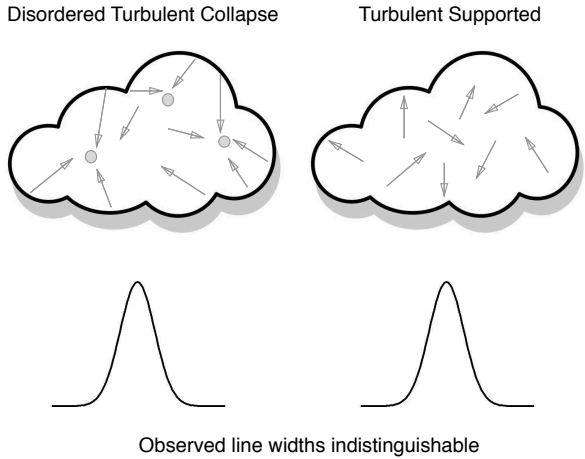}
\caption{Cartoon of the self-gravity vs. turbulence models. \textit{Top left}: Massive cloud with non-thermal motions generated only by the gravitational attraction of multiple centers of collapse. \textit{Top right}: the same cloud with motions generated by random turbulence. \textit{Bottom}: both self-gravity and turbulence generate non-thermal motions. The observed velocity dispersion (and the corresponding \ak) are the same and we cannot distinguish the origin of the non-thermal motions.}    
\label{fig:gravo-turbulent_model}
\end{figure}

In this gravo-turbulent scenario, $a_{k}$ is proportional to $a_{\mathrm{G}}$, and

\begin{equation}\label{eq:sigma_surf_dens}
\frac{a_{\mathrm{k}}}{a_{\mathrm{G}}}=\frac{\sigma^{2}}{\mathrm{R}}\frac{5}{\pi G\Sigma}=\alpha_{vir}\quad\Longrightarrow\quad a_{\mathrm{k}}=\alpha_{vir}a_{\mathrm{G}}
\end{equation}

\noindent except, in this formulation, the interpretation of  $\alpha_{vir}$ is substantially different. We cannot disentangle the turbulent from the gravitationally driven components in the observed \ak\ of each region. The measured gravo-turbulent acceleration in a region with \afor$\leq1$ could predominantly originate from either chaotic gravitational collapse or local turbulence supporting against the collapse, leading to two different star forming scenarios (see Figure \ref{fig:gravo-turbulent_model}). This ambiguity is particularly significant in GMCs and clumps. Similarly, in single cores, where the collapse is less chaotic, the contributions to \ak\ may be from the ordered motions due to the local collapse (and therefore the gravity) and from local turbulence, which are still indistinguishable. In the extreme case of all non-thermal motions being driven by self-gravity alone, $a_{\mathrm{k}}$=$a_{\mathrm{G}}$ and \avir=1. Therefore as long as the measured \avir\ is lower than 2, $a_{\mathrm{G}}$ can still account for the majority of the observed non-thermal motions.

The advantage of this formulation however is the interpretation of the dynamics in a statistically significant sample of star forming regions. If all the non-thermal motions in each region were to be gravitationally driven, then as the gravitational acceleration increases one would expect a linear increase of the observed acceleration towards higher density regions. In a more realistic context, because gravity is not the only force in play, if observing regions at increasing surface densities (i.e. increasing $a_{\mathrm{G}}$) we observe an increase of \ak, it suggests that \textit{on average} the majority of the non-thermal motions originate from gravitationally driven chaotic collapse, such that \ak\ is in fact a ``gravo-turbulent acceleration".

The Heyer relation, which shows a strong indication of increasing $\sigma/\mathrm{R}^{0.5}$ with increasing mass surface density, is equivalent to the \ak\ vs. \aG\ relationship: the Heyer relationship is indeed equivalent to Equation \ref{eq:sigma_surf_dens}, by taking the square root of \ak\ and imposing $\alpha_{vir}=1$. The Heyer relation correctly connects velocity dispersion, radius and surface density of the regions but has no direct physical interpretation of these quantities. 

The Larson relation is a specific case of Equation \ref{eq:sigma_surf_dens}. For constant \avir\ and constant $\Sigma$, Equation \ref{eq:sigma_surf_dens} implies $\sigma\propto\mathrm{R}^{0.5}$. In other words, when the gravitational acceleration is similar among different clouds, i.e. for similar values of the surface density of the clouds, and in condition of  ``virial equilibrium'' (whether  for $\alpha_{vir}=1$ or $\alpha_{vir}=2$), then the acceleration in the system is fixed.

\begin{center}
\begin{table*}
\centering
\begin{tabular}{c|c|c|c|c|c|c|c|c|c|c}
\hline
\hline
Clump & RA (J2000) & Dec (J2000) & R & $\Sigma$ & $\sigma_{nth}$  & $\mathcal{M}$ & \avir & \hco\ spec. & $\mathcal{A}_\mathrm{TG}$(\hco)  &$\mathcal{A}_\mathrm{TG}$(HNC)\\
      &   $(hh:mm:ss)$ & $(dd:mm:ss)$ & (pc) & (g cm$^{-2}$) & (km/s) &  &  & \\
  \hline     
     15.631-0.377   &   18:20:29.1 & --15:31:26  &  0.54   &   0.06 & 0.30   &   2.68   &   0.21  & Sym              &   0.11$\pm$0.01 & 0.03$\pm$0.01  \\
     18.787-0.286   &   18:26:15.3 & --12:41:33  &  0.69   &   0.27 & 1.07   &   9.82   &   0.48    & Asym           &  1.74$\pm$0.23 & 0.91$\pm$0.08 \\
     19.281-0.387   &   18:27:33.9 & --12:18:17  &  0.67   &   0.10 & 0.47   &   4.28   &   0.25   & Asym            &  0.80$\pm$0.09 & 0.40$\pm$0.04 \\
     22.53-0.192   &   18:32:59.7 & --09:20:03  &  0.80   &  0.16  & 1.25   & 11.45   &   0.92   & Asym             & 1.70$\pm$0.12 & 1.21$\pm$0.10 \\
     22.756-0.284   &   18:33:49.1 & --09:13:04  &  0.55   &   0.14 & 0.95   &   8.68   &   0.88   & Asym            &  1.80$\pm$0.77 & 0.54$\pm$0.08 \\
     23.271-0.263   &   18:34:38.0 & --08:40:45  &  0.72   &   0.13 & 0.94   &   8.65   &   0.75   & Asym            &  0.93$\pm$0.10 & 0.63$\pm$0.01 \\
     24.013+0.488   &   18:33:18.5 & --07:42:23  &  0.81   &   0.20 & 0.91   &   8.33   &   0.40   & Asym            &  1.15$\pm$0.18 & 1.07$\pm$0.15 \\
     25.609+0.228   &   18:37:10.6 & --06:23:32  &  0.97   &   0.22 &  1.05  &   9.64   &   0.40   & Asym            &  0.69$\pm$0.14 & 0.36$\pm$0.04 \\
     25.982-0.056   &   18:38:54.5 & --06:12:31  &  0.80   &   0.09 &  0.69  &   6.31   &   0.50   & Sym             &   0.27$\pm$0.01 & 0.18$\pm$0.01 \\
     28.178-0.091   &   18:43:02.7 & --04:14:52  &  0.85   &   0.19 &  1.07  &   9.79   &   0.54   & Asym            &  0.92$\pm$0.25 & 0.04$\pm$0.01 \\
     28.537-0.277   &   18:44:22.0 & --04:01:40  &  0.67   &   0.17 &  0.78  &   7.16   &   0.41   & Asym            &  0.43$\pm$0.06 & 0.38$\pm$0.03 \\
     28.792+0.141   &   18:43:08.8 & --03:36:16  &  0.61   &   0.08 &  0.99  &   9.06   &   1.55   & Sym             &   0.02$\pm$0.01 & 0.00$\pm$0.00 \\
     30.357-0.837   &   18:49:40.6 & --02:39:45  &  0.67   &   0.06 &  0.57  &   5.19   &   0.68   & Sym            &  0.14$\pm$0.06 & 0.28$\pm$0.07 \\
     31.946+0.076   &   18:49:22.2 & --00:50:32  &  0.82   &   0.14 &  1.19  & 10.87   &   0.94   & Asym             & 2.66$\pm$0.27 & 0.43$\pm$0.03 \\
     32.006-0.51   &   18:51:34.1 & --01:03:24  &  0.70   &   0.06 &  0.31  &   2.85   &   0.18   & Sym              & 0.20$\pm$0.02 & 0.04$\pm$0.01 \\
     34.131+0.075   &   18:53:21.5 & +01:06:14  &  0.55   &  0.11 &  0.74  &   6.74   &   0.72   & Asym              &  0.51$\pm$0.16 & 0.21$\pm$0.04 \\
\hline
\end{tabular}
\caption{Clump properties. Col. 1: name of the clumps as defined PI; Cols. 2-3: Coordinates of the sources, defined as the peak identified in the dust 250 \mum\ continuum maps; Col. 4: radius of the clumps as defined by the dust continuum emission; Col.5: mass surface density derived from the FIR/sub-mm dust continuum emission; Col. 6: non-thermal component of the 1D velocity dispersion. Col.7: Mach number; Col. 8: virial parameter. Col. 9: keyword to identify if the clump \hco\ spectrum shows a symmetric or asymmetric profile based on the definition of the asymmetry parameter $\mathcal{A}_{\mathrm{TG}}$ as discussed in the text. Col. 10 and 11: Measured $\mathcal{A}_{\mathrm{TG}}$ for both \hco\ and HNC spectra respectively.}
\label{tab:virial_params}
\end{table*}
\end{center}

\subsection{Implications for 70\mum\ quiet clumps}\label{sec:surf_dens_threshold}

\begin{figure} 
\centering 
\includegraphics[width=9.5cm]{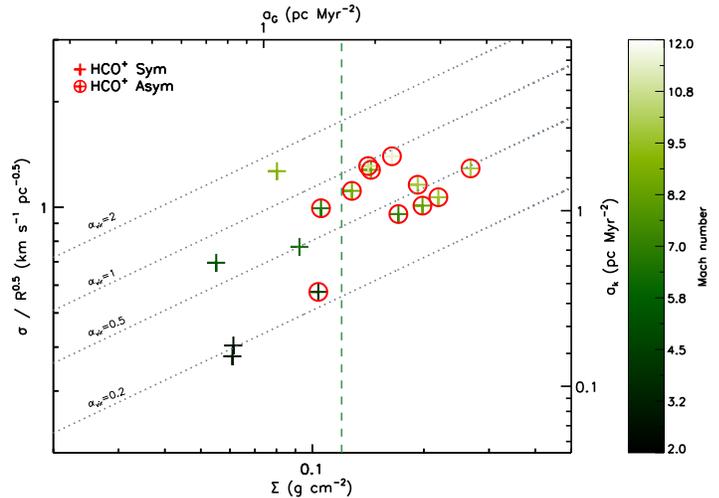} 
\caption{Heyer diagram for our 70\mum\ quiet clumps. Clumps are colour-coded with respect to the value of the respective Mach number. Also on this and the following figures we show axes in acceleration units. Clumps for which the \hco\ spectrum is asymmetric ($\mathcal{A}_\mathrm{TG}>0.4$) are shown as circled plus signs. All of the clumps with \aG$\geq1.6$ pc Myr$^{-2}\cong0.12$ g cm$^{-2}$ (the green-dotted line) show signs of dynamical activity. The grey-dotted lines follow the loci of points for constant values of the virial parameter.}  \label{fig:Sigma_v_radius_infall}
\end{figure}

Figure~\ref{fig:Sigma_v_radius_infall} shows the Heyer relation in the same units of the original H09 work, $\sigma$/R$^{0.5}$ and $\Sigma$, and in the $a_{\mathrm{k}}$ and $a_{\mathrm{G}}$ units. The clumps which show asymmetric \hco\ spectra are indicated with red circles. The clumps follow a $a_{\mathrm{k}}$ vs. $a_{\mathrm{G}}$ relationship and, in agreement with the discussion in Section \ref{sec:virial}, we interpret this trend as an indication that, on average, the non-thermal motions in these clumps may be mostly driven by self-gravity itself.

From Figure~\ref{fig:Sigma_v_radius_infall} we can see that all the clumps with $\Sigma\geq0.12$ g cm$^{-2}$ show \hco\ line asymmetries as defined in Section \ref{sec:clump_dynamics}. All our clumps are also dominated by supersonic, non-thermal motions (see Table \ref{tab:virial_params}). The average Mach number $\overline{\mathcal{M}}$ of the clumps with $\Sigma\geq\Sigma_{t}$ is $\simeq67\%$ higher on average than the corresponding Mach number of the less dense clumps. 
These results also suggest that $\Sigma_{t}=0.12$ g cm$^{-2}$ could be seen as a threshold above which clumps show more evident signs of dynamical activity, perhaps indicative of gravitational collapse at clump scales.

\section{Gravitationally driven \ak\ at all scales?}\label{sec:gravitationally_driven}

\subsection{From GMCs to cores}\label{sec:Heyer}

To explore the kinematics of the clumps in the context of both their environment (GMCs) and sub-structures (clumps and cores),  in Figure \ref{fig:Sigma_v_radius} we plot the $a_{\mathrm{k}}$ vs. $a_{\mathrm{G}}$ relation (top panel) and the dependency of the virial parameter with the surface density (bottom panel). We consider data from the literature for three different loci of points (GMCs, massive clumps and massive cores at different evolutionary stages), as follows:

\begin{figure*}
\centering
\includegraphics[width=15cm]{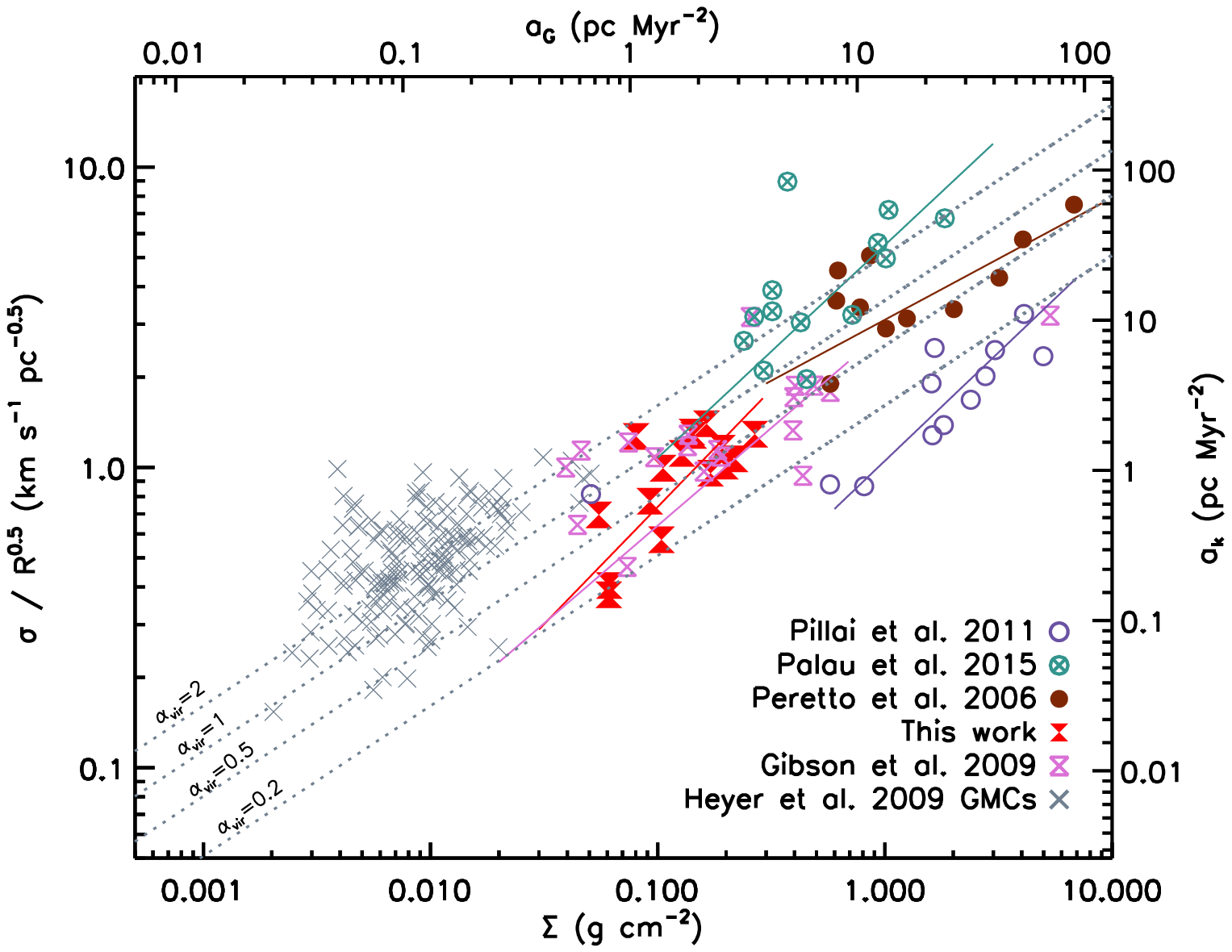}
\includegraphics[width=15cm]{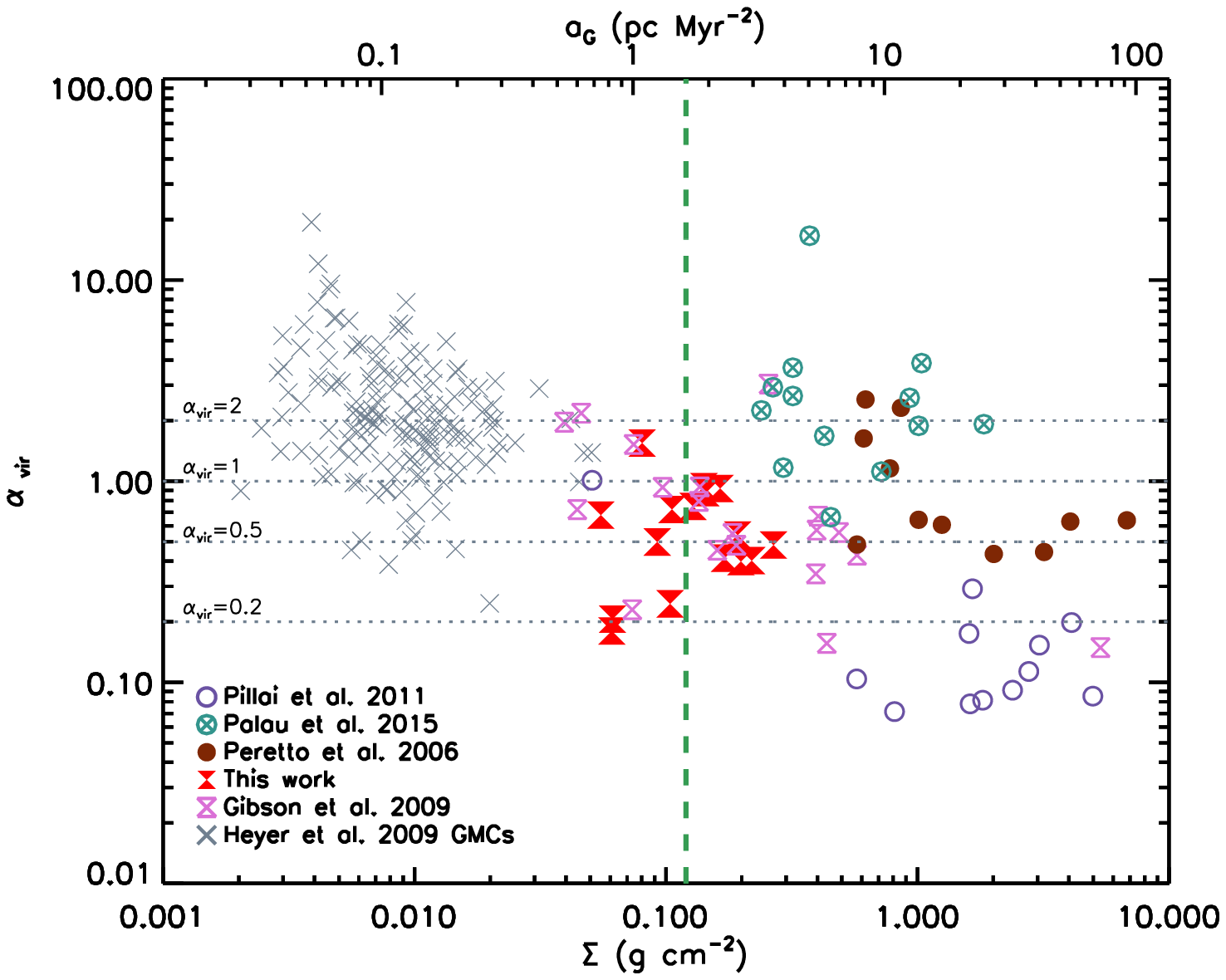}
\caption{\textit{Top}: $a_{\mathrm{k}}$ vs.$a_{\mathrm{G}}$ relation described in Equation \ref{eq:sigma_surf_dens}. The red hourglass signs are our clumps. The grey crosses are the GMC values described in H09. The light purple hourglasses are the massive clumps identified by \citet{Gibson09}. The brown-filled circles are the pre- and proto-stellar cores identified in NGC2264 by \citet{Peretto06}. The cyan-crossed circles are the massive protostellar cores described in \citet{Palau15}. Finally, the blue circles are massive cores observed by \citet{Pillai11}. The dotted lines show constant values of the virial parameter. \textit{Bottom}: Virial parameter as a function of the gravitational acceleration for the same objects as in the top panel. The green dotted line delimits the $\Sigma_{t}\geq0.12$ g\,cm$^{-2}$ region. The grey dotted lines show constant values of \avir.}    
\label{fig:Sigma_v_radius}
\label{fig:virial_a_G}
\end{figure*}

\begin{itemize}
\item[1.] The large, relatively low density GMCs traced with $^{13}$CO occupy the left-hand side of the diagram. The points show the GMC data of the cloud as a whole (size 10-100 pc) discussed in H09. The clouds span a range of masses and surface densities of $2\lesssim\mathrm{M}\lesssim2\times10^{6}$ M\sun\ and $2\times10^{-3}\lesssim\Sigma\lesssim5\times10^{-2}$ g cm$^{-2}$ respectively (H09).  

\item[2.] The central region of the diagram is occupied by our massive clumps and the sample of massive clumps analysed in \citet{Gibson09} (size 0.6-4.0 pc, mass $200\lesssim\mathrm{M}\lesssim2.5\times10^{3}$ M\sun\ and surface density $8\times10^{-3}\lesssim\Sigma\lesssim6$ g cm$^{-2}$) selected to be dark at 8 \mum\ and observed with CS emission line, a tracer of dense gas in star forming regions. This dataset, combined with the H09 data were used in \citet{Ballesteros-Paredes11} to discuss their global collapse model.
 
\item[3.] The right-hand side includes core-scale regions from different surveys, focusing on young prestellar and protostellar cores. The brown points are dense cores (size 0.03-0.05 pc, mass $3\lesssim\mathrm{M}\lesssim40$ M\sun\ and surface density $0.6\lesssim\Sigma\lesssim6.8$ g cm$^{-2}$) embedded in the massive clumps NGC2264 C and D observed by \citet{Peretto06}. The surface density of each core has been derived from the size and mass in Table 2 of \citet{Peretto06}. We used the data for the 11 resolved cores with a well defined core radius. These clumps are located in the Mon OB-1 molecular cloud complex at $d\simeq800$ pc and here we refer to the results from the \n2h\ $(1-0)$ data obtained with IRAM 30m. The blue circles are pre-protoclusters (size 0.04-0.4 pc, mass $20\lesssim\mathrm{M}\lesssim1.2\times10^{3}$ M\sun\ and surface density $5\times10^{-2}\lesssim\Sigma\lesssim5$ g cm$^{-2}$) observed with NH$_{2}$D in G29.96−0.02 and G35.20−1.74 (W48) by \citet{Pillai11} using the Plateau de Bure Interferometer. Finally, the cyan crossed circles are a sample of 13 evolved cores (size 0.1 pc, mass $9\lesssim\mathrm{M}\lesssim69$ M\sun\ and surface density $0.2\lesssim\Sigma\lesssim1.8$ g cm$^{-2}$) observed in \nh3\ $(1-1)$ with the Very Large Array (VLA) by \citet{Palau15}. We estimated the mass surface density assuming the core masses in Table 1 of \citet{Palau15} and assuming a fixed core radius as described in \citet{Palau15}. 

\end{itemize}

\begin{figure}
\centering
\includegraphics[width=9.5cm]{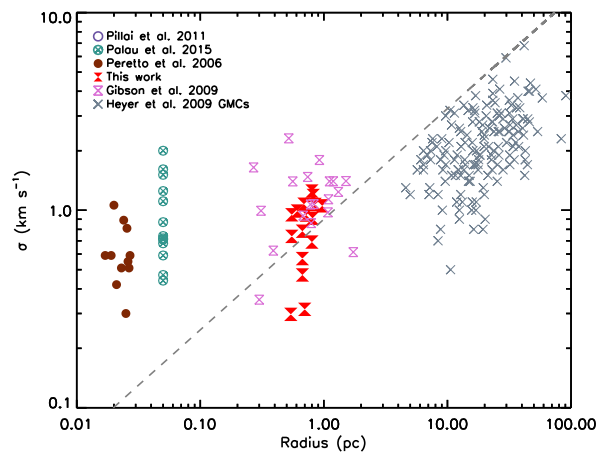}
\caption{Velocity dispersion against radius relation. Symbols are the same of Figure \ref{fig:Sigma_v_radius}. Grey-dotted line is the Larson's relation as found by \citet{Heyer04}:$\sigma\propto\mathrm{R}^{0.56}$. The cores in the survey of \citet{Palau15} have radius fixed to R=0.05 pc.}   
\label{fig:Larson}
\end{figure}

From the top panel of Fig.~\ref{fig:Sigma_v_radius} it is evident that, globally, $a_{\mathrm{k}}$ \textit{consistently} increases with increasing $a_{\mathrm{G}}$. One consequence of this result is that the first Larson's relation, $\sigma\propto\mathrm{R}^{0.5}$, is inconsistent from GMC to core scales. In Figure \ref{fig:Larson} we show the $\sigma$ against R relations for these surveys, overlaid with the \citet{Heyer04} relation $\sigma\propto\mathrm{R}^{0.56}$. The surveys together do not follow a $\sigma\propto\mathrm{R}^{0.5}$ relation, and in particular, massive clumps and cores have similar non-thermal motions at all scales in the range $0.01\lesssim\mathrm{R}\lesssim2$ pc. Also, there is no clear global trend of the virial parameters from clouds to cores. To emphasize this, in the bottom panel of Figure \ref{fig:virial_a_G} we show GMCs, clumps and cores in the \avir\ vs. $\Sigma$ plane. The green dotted vertical line indicates the $\Sigma_{t}=0.12$ g cm$^{-2}$ surface density threshold. Clouds and cloud fragments span a wide range of \avir\ at all scales and there is no evident distinction in the $\alpha_{vir}$ vs. $\Sigma$ plane below and above $\Sigma_{t}$. 

If we focus on GMCs alone, there is a large dispersion in the $a_{\mathrm{k}}$ vs. $a_{\mathrm{G}}$ plane, and the $a_{\mathrm{k}}$ and $a_{\mathrm{G}}$ relation is less obvious. Indeed, GMCs have a reasonably good Pearson's correlation coefficient (0.47) in the $\sigma$ vs. R diagram in Figure \ref{fig:Larson}. The average virial parameter for the GMCs sample is $\alpha_{\mathrm{GMC}}=1.9$ (H09) with many clouds having $\alpha_{\mathrm{GMC}}>2$ (see also Figure \ref{fig:virial_a_G}, bottom panel). This value is higher than the average value of  $\alpha_{vir}$ found in our clumps and in massive young cores as discussed in the next paragraphs. With the classical virial analysis, this would mean that the majority of the clouds are unbound on large scales. A possible explanation could be that the masses of GMCs are underestimated by a factor of 2-3 due to the LTE approximation and the assumption of constant CO abundance (H09). Otherwise, this could instead be due to a high level of turbulence in the clouds which could lead to their dispersal in the absence of a confining pressure due to HI envelope.

Alternatively, in the description of the virial parameter as a ratio of accelerations, the high values of \avir\ for GMCs implies that most of the non-thermal motions cannot be solely accounted for by the gravitationally driven acceleration. One possible explanation is that the external pressure acts as an additional confining force that contributes to the observed \ak\ \citep[e.g.][]{Field11}. In Section \ref{sec:external_pressure} we discuss the implications of accounting for an external pressure.

At clump scales, both surveys plotted in Figure \ref{fig:Sigma_v_radius} (top panel) show an increase of $a_{\mathrm{k}}$ with $a_{\mathrm{G}}$. Since the parameters in each survey have been estimated using different approaches, we fitted the samples of each survey separately. To do this we used the \texttt{linfitex} IDL routine accounting for errors in both the estimation of $a_{\mathrm{k}}$  and $a_{\mathrm{G}}$. We consider an error of 30$\%$ in the estimation of the mass surface density (i.e. $a_{\mathrm{G}}$) and an error of 20$\%$ in the velocity dispersion and radius (i.e. $a_{\mathrm{k}}$) estimations. The results of the best linear fit to each survey in the log-log space are shown as straight lines in Figure \ref{fig:Sigma_v_radius}, top panel. The slopes are $A_{clumps}=0.78\pm0.39$ and $A_{Gib.}=0.65\pm0.11$ for our clumps and the \citet{Gibson09} clumps respectively. Noticeably, these values are only slightly higher but still compatible with a slope of 0.5, which would correspond to the expected slope in case of constant value of the virial parameter going towards regions of higher surface density, i.e. a linear increase of $a_{\mathrm{k}}$ at increasing $a_{\mathrm{G}}$.

At core scales, the samples of \citet{Pillai11} and \citet{Peretto06}, tracing massive prestellar and protostellar cores lie mostly below \avir=1, which is similar to what we find for our clumps (Figure \ref{fig:virial_a_G}). The fits to these samples are also shown on the top panel of that Figure, with slopes of $A_{Pil.}=0.72\pm0.09$  and $A_{Per.}=0.41\pm0.11$ respectively, broadly consistent with an increasing gravo-turbulent acceleration towards regions of higher surface densities. The sample of more evolved cores from \citet{Palau15} lies close to or above $\alpha_{vir}=2$. Although this means that most of the non-thermal motions cannot be accounted for by gravity alone, we still observe an increase of \ak\ with \aG with a slope of $A_{Pal.}=0.71\pm0.17$, similar to what is observed in the samples of younger cores. The increased linewidth in these more evolved cores could perhaps be due to the local injection of turbulence from outflows \citep{Sanchez-Monge13,Palau15}.  

A caveat of this analysis comes from the fact that each survey has its own systematics and methodology, making a direct comparison between surveys difficult. For instance, different tracers may look at distinct regions along the line of sight and the estimated value of the linewidth (and therefore \avir) for clouds or cloud fragments in each survey may be significantly affected by the chosen tracer \citep[e.g.][]{Palau15}.

Furthermore, in the analysis above, we have only considered the contribution of the gravitational acceleration to the total motions of a given region. Other forces can contribute to the non-thermal motions we observe, such as external pressure or magnetic fields. We explore the effect of these forces in Sections \ref{sec:external_pressure} and \ref{sec:magnetic_field} respectively.

\subsection{External pressure}\label{sec:external_pressure}
In galaxy-scale simulations and observations of nearby galaxies, GMCs have been found to have high values of \avir\ which would suggest clouds are not gravitationally bound at tens of parsec scales \citep{Dobbs11,Duarte-Cabral16}. It has been suggested that these clouds could be under the effect of the external ram pressure from galactic motions and the thermal pressure of the surrounding hot ISM (e.g. Duarte-Cabral et al. 2017, in prep.), an idea consistent with the observations of nearby galaxies \citep{Hughes13}. 

If clouds are under an external pressure, Equation \ref{eq:sigma_surf_dens} must be modified to account for the pressure contribution to the observed non-thermal motion. In the standard virial analysis the pressure term can be added to Equation \ref{eq:sigma_surf_dens} as \citep{Field11}:

\begin{equation}\label{eq:pressure}
\frac{\sigma^2}{\mathrm{R}}=\alpha_{\mathrm{vir}}\frac{\pi G\Sigma}{5} + \frac{4P_{ext}}{3\Sigma}.
\end{equation}

In the acceleration formulation Equation \ref{eq:pressure} becomes:

\begin{equation}\label{eq:pressure_modified_avir}
a_{\mathrm{k}}=\alpha_{vir}\ a_{\mathrm{G}} + \frac{4\pi G}{15}\frac{P_{ext}}{a_{\mathrm{G}}}\equiv\widehat{\alpha}_{vir}(a_{\mathrm{G}})\ a_{\mathrm{G}} 
\end{equation}
with

\begin{equation}\label{eq:virial_params_pressure}
\widehat{\alpha}_{vir}(a_{\mathrm{G}})= \alpha_{vir}\bigg(1+\frac{4\pi G}{15\alpha_{vir}}\frac{P_{ext}}{a^{2}_{\mathrm{G}}} \bigg).
\end{equation}

Equation \ref{eq:pressure_modified_avir} implies that \ak\ will not be proportional to only the gravitational acceleration where the external pressure dominates and when the gravitational acceleration is sufficiently low. Equation \ref{eq:virial_params_pressure} shows that the virial parameter $\widehat{\alpha}_{vir}$ of a pressure confined cloud (or cloud fragment) is greater than \avir. For large values of $a_{\mathrm{G}}$ the terms $\propto P_{ext}/a^{2}_{\mathrm{G}}$ in Equation \ref{eq:pressure_modified_avir} goes rapidly to zero and $\widehat{\alpha}_{vir}\longrightarrow\alpha_{vir}$.

The theoretical value of the external pressure generated by the neutral ISM and required to confine a molecular cloud is of the order of $P_{ext}/k\simeq10^{4}$ K cm$^{-3}$ \citep{Elmegreen89}. Observations of individual regions show a range of values, from $P_{ext}/k=5\times10^{4}$ K cm$^{-3}$ \citep{Bertoldi92} to $P_{ext}/k=5\times10^{5}$ K cm$^{-3}$ in nearby starless cores \citep{Belloche11}.

\begin{figure*}
\centering
\includegraphics[width=14cm]{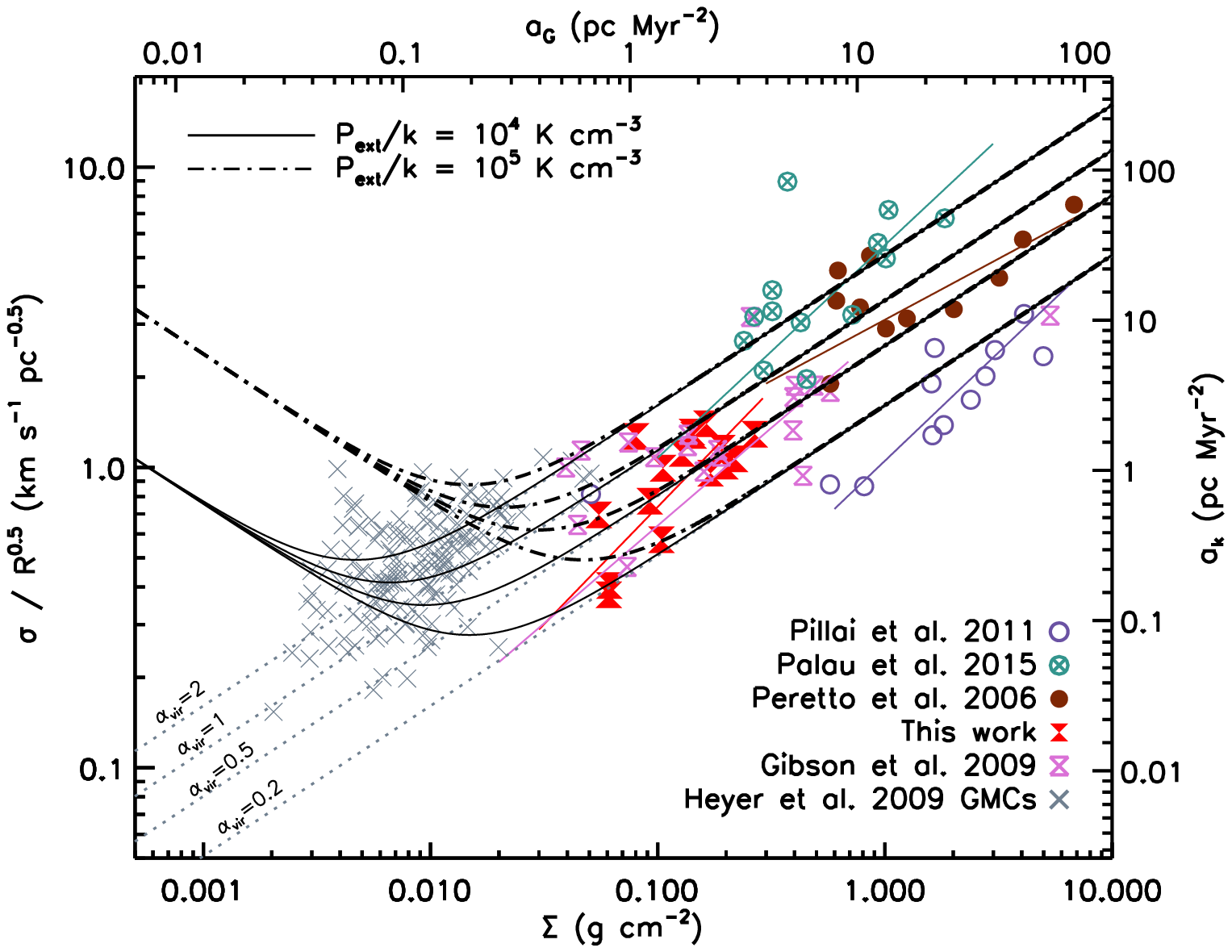}
\includegraphics[width=14cm]{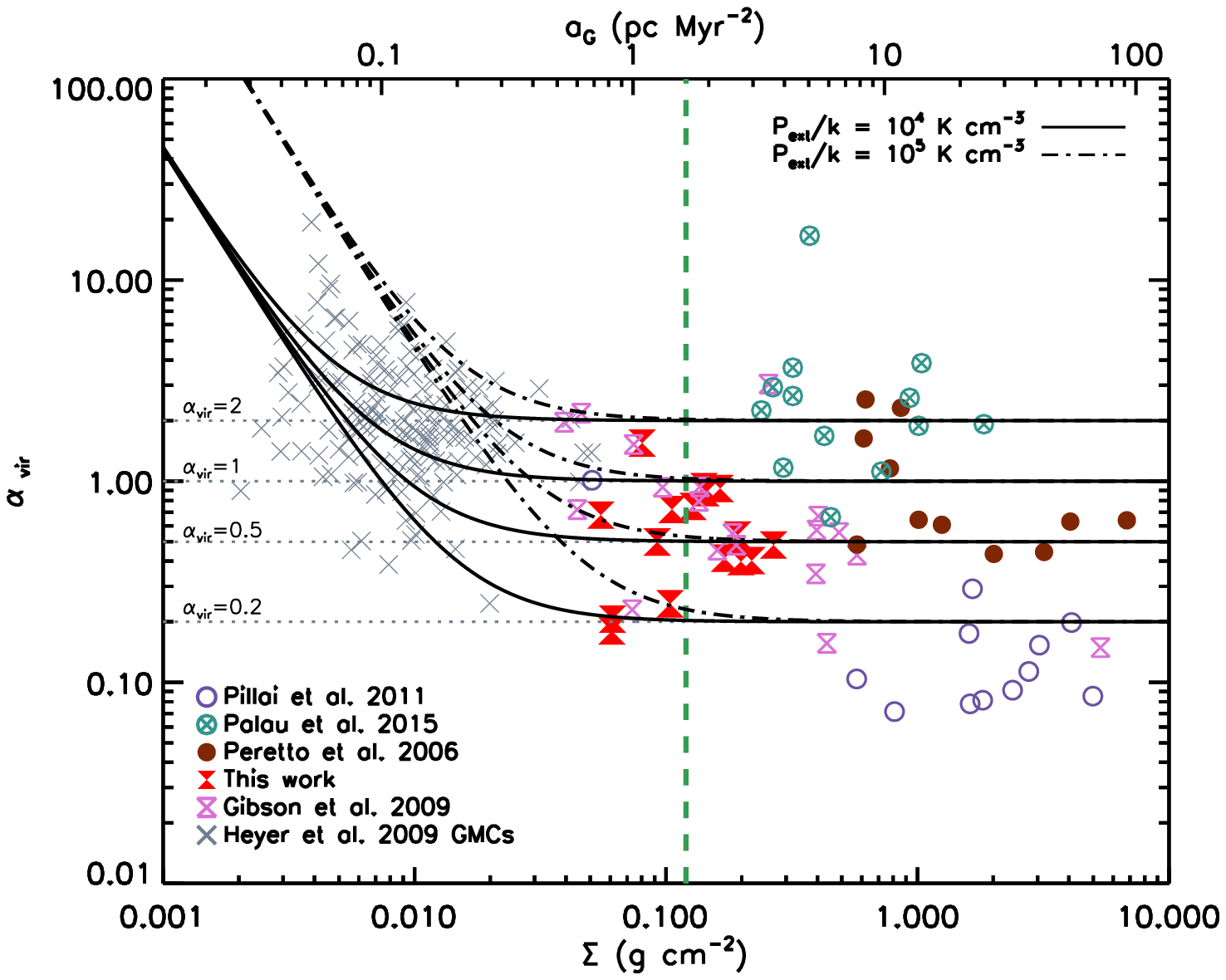}
\caption{Same of Figure \ref{fig:Sigma_v_radius}, including the solutions of Equation \ref{eq:pressure_modified_avir} for two different values of the external pressure and for fixed values of \afor.}   
\label{fig:forces_with_pressure}
\end{figure*}

In Figure \ref{fig:forces_with_pressure} we show the same data points of Figure \ref{fig:Sigma_v_radius}, this time including the loci of points occupied by the solution of Equation \ref{eq:pressure_modified_avir} for two different values of external pressure, $P_{ext}/k=10^{4}$ and $P_{ext}/k=10^{5}$ K cm$^{-3}$, (the solid and the dot-dashed line respectively) and for various values of \avir. The figure shows that with a contribution of an external pressure of e.g. $P_{ext}/k=10^{4}$ K cm$^{-3}$ at $a_{\mathrm{G}}\lesssim0.65$ pc Myr$^{-2}$, \ak\ is mostly driven by the external pressure. For a cloud with $\Sigma=0.01$ g cm$^{-2}\equiv a_{\mathrm{G}}=0.13$ pc Myr$^{-2}$ and, in absence of external pressure, \avir=0.5 the corresponding acceleration will be \ak$\simeq0.066$ pc Myr$^{-2}$. If we include a contribution of an external pressure of $P_{ext}/k=10^{5}$ K cm$^{-3}$ we will measure \ak$\simeq0.67$ pc Myr$^{-2}$ and $\widehat{\alpha}_{vir}\simeq4.9$. The high values of the virial parameter observed in GMCs can be therefore explained by motions induced by an external pressure. At $a_{\mathrm{G}}\geq0.65$ pc Myr$^{-2}$ gravity starts to dominate and the kinetic acceleration increases again with increasing gravitational acceleration at all scales, from GMCs down to cores, consistent with the threshold discussed in \citet{McKee10} and \citet{Tan14}.

We conclude that with typical values of external pressure in the ISM, most of the observed \ak\ in GMCs can be driven by external pressure, with a negligible contribution of the gravitational acceleration. However, at some point, the gas surface densities will increase to high enough values such that gravity can take over at driving the majority of the non-thermal motions, effectively acting as a gravo-turbulent acceleration.

\subsection{Magnetic fields}\label{sec:magnetic_field}
Magnetic fields can also act as a large scale support against gravity \citep{Bertoldi92,Tan13}. 

In the classical virial analysis, for the magnetic fields to act as support against gravitational collapse of a hydrostatic isothermal sphere, we obtain \citep{Kauffmann13}:

\begin{equation}\label{eq:magnetic}
B_{eq}=81 \mu\mathrm{G}\frac{M_{\Phi}}{M_{\mathrm{BE}}}\bigg(\frac{\sigma_{\mathrm{N_{2}H^{+}}}}{\mathrm{km s^{-1}}}\bigg)^{2}\bigg(\frac{R}{\mathrm{pc}}\bigg)^{-1}
\end{equation}
where $M_{\Phi}$ is the magnetic flux mass for a field of mean strength $\langle B_{eq}\rangle$ \citep{Tomisaka88}, $M_{\mathrm{BE}}$ the Bonnor-Ebert mass and $M_{\Phi}/M_{\mathrm{BE}}\simeq2/\alpha_{vir}-1$ \citep{Kauffmann13}. 

Observationally, \citet{Crutcher12} suggested, however, that there is an upper limit of the intensity of the magnetic field $B_{up}$ in a given region which depends on the gas number density, as $B_{up}\simeq150\ \mu\mathrm{G}\ (n_{\mathrm{H_{2}}}/10^{4}$ cm$^{-3})^{0.65}$. If we assume a spherical geometry and a density equal to the mean density, $B_{up}$ can be rewritten as \citep{Kauffmann13}:

\begin{equation}\label{eq:Crutcher}
B_{up}\simeq336\ \mu\mathrm{G}\ \bigg(\frac{M}{10M_{\odot}}\bigg)^{0.65}\bigg(\frac{R}{0.1\ \mathrm{pc}}\bigg)^{-1.95}
\end{equation}

\noindent We can then compare the strength of the magnetic fields required to reach the virial equilibrium (that is, $B=B_{eq}$) with the critical value $B_{up}$. 
In Table \ref{tab:magnetic_fields} we report $B_{eq}$, $B_{up}$ and the ratio between the two for our sample, where we separate the clumps above and below the surface density threshold for clarity.

\begin{figure}
\centering
\includegraphics[width=7cm]{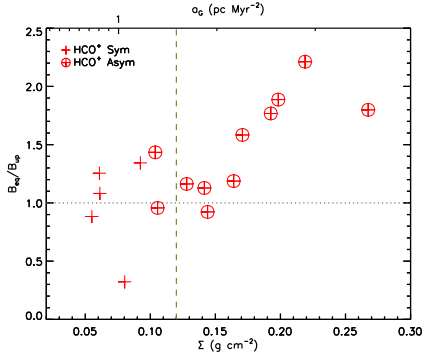}
\caption{Ratio $B_{eq}$/$B_{up}$ as function of the gravitational acceleration (i.e. surface density) of each clump. The correlation is significant with a Pearson's coefficient of 0.69. All but four clumps have $B_{eq}$/$B_{up}\geq1$ (the dotted horizontal line), i.e. they need an intensity of the magnetic field higher then the maximum allowed as predicted by \citet{Crutcher12}. The green-dashed vertical line marks the value $\Sigma_{t}=0.12$ g cm$^{-2}$.}   
\label{fig:magnetic_fields_surface_density}
\end{figure}

In Figure \ref{fig:magnetic_fields_surface_density} we show the ratio $B_{eq}$/$B_{up}$ as a function of $\Sigma$. There is a good correlation between $B_{eq}$/$B_{up}$ and $\Sigma$ (Pearson's coefficient $\rho$=0.69, which implies a probability of a non-correlation of only 0.003). This implies that for larger $\Sigma$, the magnetic fields needed to prevent collapse is increasingly greater than the maximum values predicted by \citet{Crutcher12}, and therefore they cannot halt collapse. 

In the framework of accelerations, we can instead look at how the magnetic fields could generate a negative acceleration that opposes gravity, preventing the collapse. If we take $B_{up}$ from Equation \ref{eq:Crutcher} as an upper limit for the strength of the magnetic fields in our clumps, we can estimate the maximum magnetic pressure as $P_{B_{up}}=B_{up}^{2}/(8\pi)$ and therefore derive the maximum magnetic acceleration, $a_{B_{up}}$, as

\begin{equation}
a_{B_{up}}=\frac{F_{B_{up}}}{\mathrm{M}}=\frac{P_{B_{up}}\pi\mathrm{R}^{2}}{\mathrm{M}} \qquad \rightarrow  \qquad a_{B_{up}}=\frac{B_{up}^{2}}{8\pi\Sigma}
\end{equation}

\begin{figure}
\centering
\includegraphics[width=7cm]{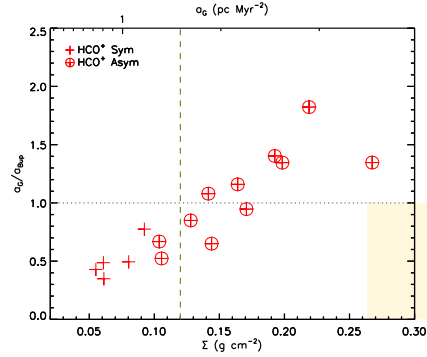}
\caption{Same of Figure \ref{fig:magnetic_fields_surface_density}, but for the ratio of the gravitational acceleration \aG\ and the maximum acceleration produced by magnetic fields $a_{B}$. The black-dotted horizontal line is in correspondence of $a_{\mathrm{G}}$/$a_{up}=1$ The green-dashed vertical line marks the value $\Sigma_{t}=0.12$ g cm$^{-2}$.}    
\label{fig:magnetic_fields_acceleration}
\end{figure}

In Figure \ref{fig:magnetic_fields_acceleration} we compare $a_{B_{up}}$ with the acceleration imposed by the gravitational potential of the clumps, \aG. The figure shows that, above a surface density of $\Sigma\simeq0.18$ g cm$^{-2}$, the maximum acceleration generated by the magnetic fields is not sufficient to overcome \aG\ and therefore the collapse will proceed. Nevertheless, we observe signs of dynamical activity in clumps with $\Sigma<0.18$ g cm$^{-2}$, even though they have \aG/$a_{B_{up}}<1$.

This could be due to the fact that $a_{B_{up}}$ is a strict upper limit to the magnetic acceleration. In fact, not only it assumes the maximum value of the magnetic fields from \citet{Crutcher12}, but it also considers the geometry of the magnetic fields such that all the gas would feel the same resistance due to the magnetic force. For instance, in a case of a uniform magnetic field threading the clump, only the gas trying to collapse perpendicular to the field lines would feel this maximum tension.

\begin{center}
\begin{table}
\centering
\begin{tabular}{c|c|c|c|c}
\hline
\hline
Clump & Type & B$_{eq}$ & B$_{up}$ & $\frac{B_{eq}}{B_{up}}$  \\
       &  & ($\mu$G) &  ($\mu$G) &   \\
\hline

18.787-0.286  &	 Asym	&	337 &   211 &    1.6  \\
19.281-0.387  & Asym		&	186 &   129 &    1.4 \\ 
22.53-0.192  & Asym		&	197 &   161 &    1.2 \\ 
22.756-0.284  &	 Asym	& 150 &   175 &    0.9	 \\
23.271-0.263  &	 Asym		&	189 &   151 &    1.2 \\ 
24.013+0.488  & Asym	&	342 &   177 &    1.9 \\ 
25.609+0.228  & Asym	&	260 &   141 &    1.8 \\ 
28.178-0.091  & Asym	&	376 &   188 &    2.0 \\ 
28.537-0.277  & Asym	&	324 &   195 &    1.7 \\
31.946+0.076  & Asym	&	158 &   140 &    1.1 \\     
34.131+0.075  & Asym		&		124 &   141 &    0.9 \\ 

& & & & \\         
          
15.631-0.377  & Sym		&	108 &   100 &    1.1 \\ 
25.982-0.056  & Sym		&	145 &   108 &    1.3 \\ 
28.792+0.141  &	 Sym		&	40 &   117 &    0.3 \\ 
30.357-0.837  & Sym     &  104 &    93 &    1.1 \\
32.006-0.51  &	 Sym		&	120 &    92 &    1.3  \\ 
 
\hline
\end{tabular}
\caption{Magnetic fields estimates  for clumps above and below the surface density threshold $\Sigma=0.12$ g cm$^{-2}$. Col. 1: name of the clumps; Col. 2: keyword to identify clumps which show symmetric (Sym) or asymmetric (Asym) \hco\ spectra. Col. 3: Intensity of the magnetic field required to reach BE equilibrium, B$_{eq}$. Col. 4: Maximum intensity of the magnetic field as estimated following \citet{Crutcher12}, B$_{up}$. Ratio between B$_{eq}$  and B$_{up}$.}  
\label{tab:magnetic_fields}
\end{table}
\end{center}

\section{Conclusions}\label{sec:conclusions}
We have used new IRAM 30m observations of a sample of 16 70\mum\ quiet clumps identified in IRDCs in the Hi-GAL survey to explore the kinematics of these regions. The clumps have been selected to be ``quiescent'', i.e. dark or very faint at 70 \mum\ and with a L/M$<0.3$ (PI). With these data we show that there is a correlation between the asymmetry of the \hco\ line profile, tracing the dynamics of the regions, and the clump surface density, with the highest surface density clumps having the most asymmetric lines, and so being the most dynamically active. 

Looking at the relationship between column density, size and velocity dispersion we demonstrate that the Heyer relation can be re-interpreted as a direct consequence of a gravo-turbulent description of the non-thermal motions in collapsing clouds and cloud fragments. In this formalism the virial parameter is described as ratio between the gravitational acceleration \aG\ and the observed acceleration \ak, the former defined as function of the mass surface density and the latter as function of the velocity dispersion and the radius of the cloud or cloud fragment.

We have used our sample of 16 massive 70\mum\ quiet clumps to explore this formalism, together with other surveys of GMCs, clumps and pre- and proto-stellar cores from the literature. 

We can summarise our findings as follows:

\begin{itemize}
\item[1.] The non-thermal motions observed in clouds and cloud fragments originate from both self-gravity and turbulence that can act against gravity itself. The two components cannot be observationally separated. However, the data show that, from the scales of clumps (and possibly GMCs) down to the cores, the global measured acceleration, $a_{\mathrm{k}}$, increases with the gravitational acceleration, $a_{\mathrm{G}}$. This suggests that, on average, the self-gravity can drive much of the observed non-thermal motions at all spatial scales. 

\item[2.] For our sample of massive 70\mum\ quiet clumps, we find asymmetric line profiles tracing dynamical activity, which we interpret as due to gravitational collapse, regardless of the estimated value of \avir\ in the single regions, suggesting that the virial parameter is not a good descriptor of the stability of a region. In our formulation, in the absence of magnetic fields and external pressure, \avir\ can instead be seen as a measure of how much the non-thermal motions can be generated by gravity. Indeed, as long as \avir$<2$ (\ak$<2$\aG), gravity can dominate the non-thermal motions. This is the case for all our clumps, which would be consistent with them all being dynamically active and undergoing collapse.


\item[3.] From our data we identify a surface density value, $\Sigma_{t}\,\simeq\,0.12$\,g\,cm$^{-2}$, above which all our clumps have highly asymmetric HNC and \hco\ spectra, which we interpret as tracing collapse. We interpret this as the threshold above which gravity is strong enough to dominate the non-thermal motions at clump scales. In a scenario in which massive protostars accrete dynamically from their parent clump, $\Sigma_{t}$ could therefore represent the minimum surface density at clump scales for high-mass star formation to occur.

\item[4.] An external pressure $P_{ext}$ which confines the GMCs such that $P_{ext}/k\simeq10^{4}-10^{5}$ K cm$^{-3}$ explains the high values of the measured non-thermal motions observed in these clouds, and does not exclude that the majority of them are globally collapsing. The contribution of the external pressure can be incorporated in the formalism of the gravo-turbulent mechanism, modifying the definition of \avir.

\item[5.] Magnetic fields, if present, can be a support against the collapse. From the classical view, we show that magnetic fields stronger than the maximum intensity of the magnetic field suggested by \citet{Crutcher12} would be required to support the majority of the clumps. A similar conclusion can be drawn by comparing the gravitational acceleration to the maximum acceleration produced by the magnetic fields.

\end{itemize}

The gravo-turbulent formulation and the $a_{\mathrm{k}}$ vs. $a_{\mathrm{G}}$ relation, if confirmed at all scales, may help our understanding of the massive star formation mechanism. If the global collapse mechanism with the gravity driving much of the non-thermal motions is correct, and the strength of the magnetic fields is not sufficient, we may for example expect to observe pure thermal Jeans fragmentation in most of the clumps. This hypothesis can be explored using high-resolution instruments such as NOEMA or ALMA.

\section*{acknowledgements}
This work has benefited from research funding from the European Community's Seventh Framework Programme. AT and GAF are supported by STFC consolidated grant ST/L000768/1 to JBCA. RJS gratefully acknowledges support from the STFC through an Ernest Rutherford Fellowship. AT wants to thank J. Ballesteros-Paredes and E. Vazquez-Semadeni for their inspiring works and S. Camera for the stimulating conversations at the origin of this work.

\bibliographystyle{mn2e}
\bibliography{gravo_turbulence_acc.blb}

\end{document}